\documentclass[10pt, conference, compsocconf]{IEEEtran}
\ifCLASSINFOpdf
  \usepackage[pdftex]{graphicx}
  \else
 \fi

\usepackage[colorinlistoftodos]{todonotes}
\usepackage[cmex10]{amsmath}
\usepackage{url}
\usepackage{hyperref}
\usepackage{listings}
\begin{document}

\title{Accelerating Radiation Therapy Dose Calculation with Nvidia GPUs}



\author{\IEEEauthorblockN{Felix Liu$^{1,2}$, Niclas Jansson$^1$, Artur Podobas$^1$, Albin Fredriksson$^2$, Stefano Markidis$^1$}
\IEEEauthorblockA{
$^1$ KTH Royal Institute of Technology, Stockholm, Sweden\\
$^2$ RaySearch Laboratories, Stockholm, Sweden\\
}
}

\maketitle

\begin{abstract}

Radiation Treatment Planning (RTP) is the process of planning the appropriate external beam radiotherapy to combat cancer in human patients. RTP is a complex and compute-intensive task, which often takes a long time (several hours) to compute. Reducing this time allows for higher productivity at clinics and more sophisticated treatment planning, which can materialize in better treatments. The state-of-the-art in medical facilities uses general-purpose processors (CPUs) to perform many steps in the RTP process.
In this paper, we explore the use of accelerators to reduce RTP calculating time. We focus on the step that calculates the dose using the Graphics Processing Unit (GPU), which we believe is an excellent candidate for this computation type. Next, we create a highly optimized implementation for a custom Sparse Matrix-Vector Multiplication (SpMV) that operates on numerical formats unavailable in state-of-the-art SpMV libraries (e.g., Ginkgo and cuSPARSE). We show that our implementation is several times faster than the baseline (up-to 4x) and has a higher operational intensity than similar (but different) versions such as Ginkgo and cuSPARSE.





\end{abstract}


\IEEEpeerreviewmaketitle

\section{Introduction}
Radiation Therapy (RT) is one of the most commonly found cancer treatments and uses various types of highly concentrated beams to kill identified cancer cells. Today roughly 50\% of patients diagnosed with cancer will receive RT sometime during the course of their illness~\cite{baskar2012cancer}. Before starting RT with a patient, a radiation therapy plan (called \textit{treatment plan} or simply \textit{plan}) must be developed by Radiation Therapy Planning (RTP). The RTP is performed by a group of medical experts that primarily rely on (and are assisted by) the result of computer software. Said software is often very complex, compute-intensive~\cite{jia2014gpu}, and -- perhaps most importantly -- very time-consuming (computing a plan can take several hours or even days), which restricts the number of plans that can be computed within a given time frame.

In today’s state-of-the-art paradigm for RTP, mathematical optimization is used to determine the settings of the treatment machine (typically a linear accelerator or a particle accelerator) that will result in as beneficial a dose to the patient as possible. The main computational bottlenecks of RTP are two: the radiation dose computation and the optimization step. Dose distributions from multiple beams, possibly under various realizations of uncertainties, must be computed in each iteration of an optimization procedure, which moreover requires the calculation of function values and often gradients to update the beam variables. In this work, we focus on the radiation dose computation.

RTP may be seen as a joint work of a radiation oncologist and a medical physicist. However, there is a critical contribution to algorithmic and software development by computer scientists, mathematicians, and physicists. An oncologist makes the first step in today’s RTP by marking contours of organs and regions of suspected malignant cancer cells that have been delineated on computed tomography (CT) images. These contours form input to the software that produces a treatment plan under the supervision of a medical physicist, based on the prescribed target dose and the tolerance doses to each organ at risk as specified in the clinical protocol or by the oncologist. The plan includes the set of operations for the treatment machine to deliver the treatment. To generate the plan, the medical physicist formulates the treatment goals as an optimization problem, which is solved by the planning software. The objective function of the problem measures the quality of the simulated treatment dose. Typically, the objective function will promote a high, uniform dose in the target volume and low dose in the surrounding organs. The optimization problems are of varying complexity, sometimes taking into account aspects of uncertainty, such as patient positioning errors, changes of the patient anatomy between treatment sessions, and calculation uncertainties. There is a trade-off between the quality of the solution and the complexity of the optimization problem. The optimization problem is typically solved iteratively, and requires calculation of the simulated treatment dose in each iteration.

A crucial aspect when it comes to the performance of the RTP is the significant advances within high-performance computing over recent years and the advent of accelerators. GPUs especially have seen successful use in accelerating many of the workloads in RTP~\cite{jia2014gpu}. High-Performance Computing systems consist of very different technologies that can lead to a considerable performance and efficiency improvement of the applications. Today, supercomputers not only feature conventional CPUs and memory technologies, e.g., DRAM. They also include computing units that can accelerate the computation and new memory technologies to speed-up the data movement.

Today, accelerators such as GPUs are often providing a majority of computing in several of the fastest computers in the world\footnote{Three out of five of the top5 systems includes GPUs (www.top500.org)} and even FPGAs are slowly gaining traction as alternatives~\cite{podobas2017evaluating}.

An example of a modern accelerator is the newly released Nvidia Ampere GPU~\cite{choquette2020nvidia}, providing a computational power of roughly 9.4 TFlop/s in double precision in one chip. However, the performance improvement comes with increased difficulty in programming the accelerators and with a change of the algorithms to better map to the underlying hardware architecture. This work addresses this challenge of programming radiation therapy dose calculation on GPUs.

The main contributions of the paper are the following:
\begin{enumerate}
\item We design and develop a new kernel for the radiation dose calculation on Nvidia GPUs. The radiation dose calculations are intended for the treatment planning system RayStation. We convert the RayStation compressed sparse matrix format to CSR and implement CSR sparse matrix vector multiplication (SpMV) for radiation dose calculation with mixed half and double precision and using CUDA cooperative groups.
\item We evaluate the performance of radiation dose calculations on different GPU systems, including a machine with Nvidia A100, and compare its performance with the performance of the state-of-the-art implementations, such as Ginkgo~\cite{anzt2020ginkgo} and cuSPARSE~\cite{naumov2010cusparse}. We find that our implementation is up to 4$\times$ faster than a baseline implementation mimicking RayStation radiation dose calculation on GPU. Our implementation performance is also comparable and in some cases faster than implementations using Ginkgo and cuSPARSE.
\end{enumerate}

The paper is organized as follows. Section~\ref{background} describes how radiation dose calculations are performed in RTP systems. We present our design and implementation of radiation dose calculation on Nvidia GPUs in Section~\ref{methodology}. Section~\ref{experiment} describes the experimental set-up and the systems we use for the performance evaluations. Section~\ref{results} presents the performance results and roofline models, and Section~\ref{related_work} briefly reviews previous related works. Finally, Section~\ref{conclusions} summarizes the work and outlines future work.


\section{Background}
\label{background}

RTP aims to create a treatment plan such that a sufficient dose is delivered to the tumor while sparing surrounding healthy tissue as much as possible.  The treatment machine needs control parameters to provide the desired dose. To determine these control parameters, a treatment planning system (TPS) is used, which is a software product designed specifically for RTP. In this work, the TPS used to create the dose deposition matrices (see Section~\ref{sec:dose_calc}) is RayStation \cite{RaystationLink}, developed by RaySearch Laboratories. RayStation is written in C++ and C\#, and it also provides Python as a higher-level scripting interface.

The radiation dose calculation is one of the most computationally intensive routines in RayStation as it is performed at each iteration in the optimization solver. Currently the dose calculation for pencil beam scanning (see Section~\ref{sec:dose_calc}) in proton therapy RayStation is done partly on CPUs. In this work, we focus on porting this routine to Nvidia GPUs using CUDA and evaluating the performance benefit that RayStation can reach by offloading the radiation dose calculation for protons to GPUs.



\subsection{Radiation Dose Calculation} \label{sec:dose_calc}
The radiation dose calculation relies on a method that first constructs the \textit{dose deposition matrix} and then multiplies it to the control parameters in the form of an input vector. Calculating the dose in the patient from the values of the control parameters becomes a single matrix-vector product. The assumption of this method is that the relationship between the control parameters and the dose is linear.
Various methods for computing the dose deposition matrix exists, including Monte Carlo simulation~\cite{ma2002monte} and pencil beam approximations~\cite{ahnesjo1992pencil}. In this work, we use RayStation's Monte Carlo simulations to compute the dose deposition matrix. The Monte Carlo simulations simulate the transport of individual particles (either protons or photons) through the tissue to generate the dose deposition matrices. Numerical noise affects Monte Carlo simulations and for this reason deposition matrices will include entries for voxels where a very small dose (due to the noise of the method) is received. This can lead to an artificial increase of the non-zero values in the dose deposition matrix.

\begin{figure}
     \centering
     \includegraphics[width=\linewidth]{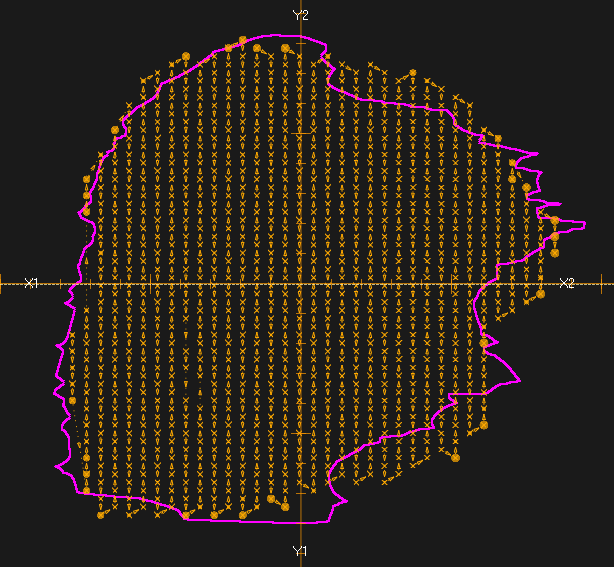}
     \caption{Illustration of the spot scanning treatment technique. The figure is shown from the perspective of the treatment beam. The purple outline shows the target volume (the tumor), and the orange crosses are the spots. The arrows between the spots indicate the direction that the beam scans the target volume.}
     \label{fig:spot_scanning}
 \end{figure}

In this work, we focus on the dose deposition matrices for patients treated with proton therapy~\cite{mohan2017proton} (other sources of radiation in radiation therapy include photons, electrons or carbon ions). The specific treatment technique used in these patient cases is \textit{pencil beam scanning}: a thin proton beam scans the target in a scanline pattern, stopping to deliver dose at a number of \textit{spots}. An illustration of the pencil beam scanning treatment technique from RayStation is shown in Figure~\ref{fig:spot_scanning}. In the figure we use the "Beam's eye view": a 2D view of the target, seen from the beam's perspective. The dose deposition matrix in this case is a mapping between the spot weights, a measure of the amount of radiation delivered by the pencil beam at each spot (the orange crosses in Figure~\ref{fig:spot_scanning}), and the resulting dose in the voxels of the patient, measured in Gray. With this in mind, a column of the dose deposition matrix is the contribution of a single spot to the dose in all voxels of the patient. Since a single spot will not contribute to the resulting dose in most voxels, \textbf{dose deposition matrices are typically highly sparse} (for instance, in the test cases we show in this paper, the percentage of the non-zero elements varies between 0.6\% to 2\%).

The characteristics of dose deposition matrices can vary significantly due to a number of factors. Firstly, different treatment modalities, such as photon and proton treatments, will result in matrices with different characteristics because the dose deposition and physics differ. Furthermore, dose deposition matrices can differ for different treatment sites, and between cases as well, simply due to differences in patient geometry, anatomy and setup.

Since the optimization problems from RTP are often complex and nonlinear, iterative methods for solving them are necessary. As such, solving the optimization problem often requires the dose to be calculated multiple times, meaning that dose calculation can become a significant bottleneck. Note also that the dose deposition matrix does not change during the optimization process; only the machine parameters (the input vector in our algorithm) do.

Thus, a fast and efficient method for computing the dose using the dose deposition matrix is required. Since the dose deposition matrices can become very large, the use of accelerators like GPUs could be used to speed up optimization times. Better performance of the optimization algorithm not only improves the productivity in clinics, but can also enable the creation of higher quality plans because it enables the planners to explore more treatment options in a given time, as well as through the use of more sophisticated and computationally demanding optimization methods. Examples of such methods include robust optimization, where uncertainties in treatment delivery due to, e.g., changes in the patient geometry between successive treatment sessions and patient movement during treatment delivery can be taken into account by the optimization algorithm.
\subsection{Dose Deposition Matrix Characteristics}

We test our radiation dose calculations on GPUs with two use cases: a liver case with four beams from different angles, and a prostate case with two parallel opposed beams. Each beam is characterized by its own dose deposition matrix. We thus test a total of six dose deposition matrices. The dimensions of each of the matrices are given in Table~\ref{tab:dimensions}, together with the number of non-zeros. Recall that the number of rows in the matrix is the number of voxels in the dose grid, and the number of columns is the number of spot weights. Because the number of dose grid voxels is typically much higher than the number of spot weights, the dose deposition matrices are highly skewed in the row direction: the number of rows is 40-200$\,\times$ the number of columns in the test cases.

\begin{figure}[t]
    \centering
    \includegraphics[width=\linewidth]{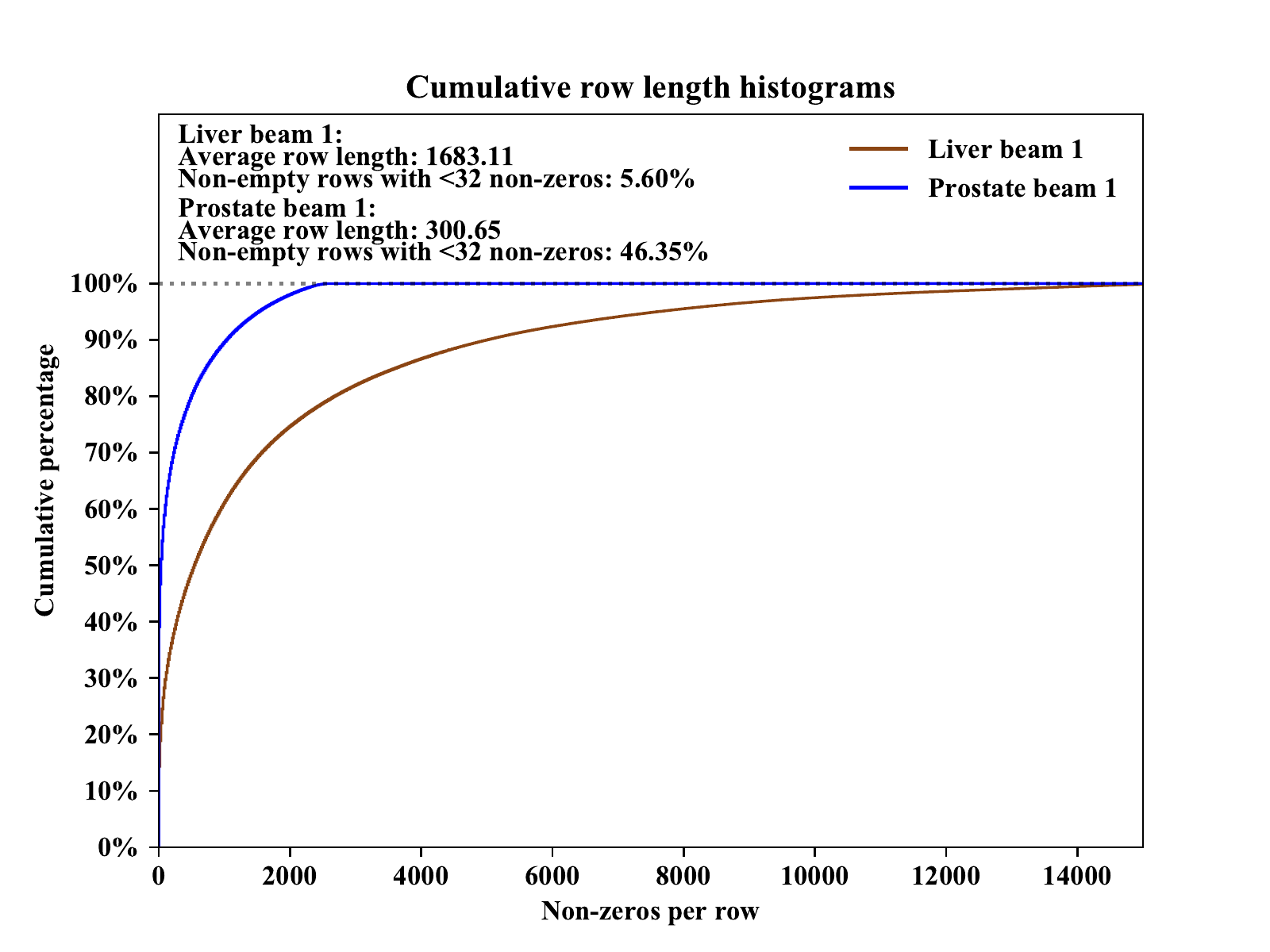}
    \caption{Cumulative row length histograms for liver~beam~1 and prostate~beam~1. The rows with length 0 are excluded in the histograms. In both liver and prostate~beam~1, ~70\% of the rows have length 0.}
    \label{fig:nonzero-row-hist}
\end{figure}

\begin{table}[t]
    \centering
    \begin{tabular}{|c|c|c|c|c|c|}
         \hline
         beam no. & Rows    & Columns & non-zeros  & non-zero & size (GB) \\
                  &         &         &            & ratio    &     \\
         \hline
           Liver 1 & 2.97e6 & 6.80e4   & 1.48e9 & 0.73\% & 8.880    \\
           Liver 2 & 2.97e6 & 6.77e4   & 1.28e9 & 0.64\% & 7.672    \\
           Liver 3 & 2.97e6 & 6.99e4   & 1.39e9 & 0.67\% & 8.368    \\
           Liver 4 & 2.97e6 & 6.32e4   & 1.84e9 & 0.98\% & 11.04    \\
           Prostate 1 & 1.03e6 & 5.09e3 & 9.50e7 & 1.81\% & 0.5744   \\
           Prostate 2 & 1.03e6 & 4.96e3 & 9.51e7 & 1.86\% & 0.5747   \\
      \hline
    \end{tabular}
    \caption{Characteristics of the dose deposition matrices used in this work.}
    \label{tab:dimensions}
\end{table}
To gain insights about the unstructured nature of the sparse matrices, we can investigate how the non-zeros are distributed throughout the matrix (specifically along the rows since we are going to parallelize the SpMV algorithm along that direction). Figure~\ref{fig:nonzero-row-hist} shows the cumulative function for the number of non-zero values per row of the dose deposition matrices for beam~1 in the liver and prostate cases. The cumulative function provides us the percentage of rows with less than a certain number of non-zero elements (x-axis). The plots presents also the average number of non-zeros per row and the percentage of the non-empty rows that have less than 32 non-zero elements. In this work, we implement an SpMV algorithm that is most efficient when there are more than 32 non-zero elements per row: from Figure~\ref{fig:nonzero-row-hist} we see that cases violating this account only for 5.6\% and 14.2\% for the liver and prostate cases, respectively.

The histogram and statistics of the row lengths indicate a high level of irregularity in the matrices. Many rows are relatively short for the liver case, with lengths in the tenths or hundreds, while other rows have around 16000 non-zeros. For the prostate case, the difference in row lengths is also significant, though not as dramatic as the liver case. This could mainly be due to the much smaller number of total columns.

\subsection{Sparse Matrix Formats}
As mentioned previously, dose deposition matrices are typically highly sparse, due to each part of the treatment beam contributing dose to only a small part of the patient volume. As such, the problem of computing the dose in the patient given a dose deposition matrix becomes an SpMV.

Sparse matrices can be represented in several formats. Some common representations include coordinate list (COO) and compressed sparse row (CSR). In COO, three arrays are used to store the matrix. One with the non-zero values, one with the row index for each value and one with the column index for each value. CSR format also uses three arrays, one for non-zero values (in row-major order), one with column indices, and one with the starting index of each row in the previous two arrays. There also exists other sparse matrix representations specifically designed for performance on GPUs and SIMD architectures. Examples of this include the ELLPACK, and SELL-C-sigma~\cite{kreutzer2014unified} formats, designed for SIMT and wide SIMD architectures.

RayStation uses a custom storage format to compress the dose deposition matrices. In this work, we convert RayStation's custom compressed format to CSR. The use of CSR storage format allows us to gain insights into the structure of dose deposition matrices from RTP and obtain insights about achievable performance on different hardware. RayStation's custom sparse matrix format was developed for CPUs at a time when the amount of available memory in most systems was much smaller than today. Investigating other storage formats, such as ELLPACK, and SELL-C-sigma, will be a topic of future work.

\subsection{RayStation Requirements}
To integrate the SpMV kernels investigated in this paper into RayStation, we need to meet two requirements.
First, \textbf{the input and output vectors} of the SpMV product \textbf{must be in double precision}. The reason for this is that lower precisions have been found to affect the numerical stability of the optimization engine used in RayStation. However, due to the size of dose deposition matrices from larger patient cases, the entries in \textbf{the dose deposition matrix should preferably be stored in lower precision}. The current CPU implementation in RayStation uses 16 bits to store the entries in the matrix. To match this level of precision, IEEE-754 half precision floating point can be used in the GPU implementations

Second, the radiation dose calculation must be reproducible on the same system: computing the matrix-vector product using the same input data and system must yield exactly the same result at each time run. The non-associativity of floating point arithmetic implies that certain methods and algorithms cannot be used. For instance, our algorithm cannot implement reductions using atomic operations, where the order of operations is not set in advance (this would lead to non-reproducible results on different runs on the same system).
\section{Methodology}
\label{methodology}
Our implementation of the SpMV for the radiation dose calculation is an adaption of the "Vector CSR-kernel" presented in the seminal paper by Bell and Garland~\cite{bell2009implementing}. The main difference is the use of CUDA cooperative groups, introduced recently in CUDA 9.0. These allow for efficient synchronization and collective operations in threads belonging to the same warp.

We show a stripped-down version of our CSR sparse matrix-vector multiplication in Listing~\ref{spmv}. To exploit CUDA cooperative groups, we assign one thread warp of 32 threads to process each row of the input matrix. The main motivation behind assigning one thread warp to each row, as opposed to just a single thread, is a more favourable memory access pattern. The reason behind this is that at each iteration of the inner loop in the SpMV kernel (see lines 27-28 in Listing 1), the threads in each warp access consecutive elements in the data array. This is not the case when assigning one thread per row in the matrix. In that case, at each iteration each thread in a warp accesses elements from different rows, which will not be consecutive in the data arrays.

Thus, each row in the matrix is processed by one thread warp, with an intra-warp reduction being performed for each row before the result is stored in the output vector. In our implementation, the coordination of threads and warps is largely handled via CUDA's cooperative groups API~\cite{CooperativeGroups}: we use a \texttt{tiled\_partition} to partition the thread blocks into chunks of 32 threads, such that each part of the tiled partition contains one thread warp. We utilize the \texttt{reduce} function from the CUDA cooperative groups API to perform the intra-warp reductions.

Several libraries, such as cuSPARSE and Ginkgo, provide optimized CSR SpMV implementations. However, in our implementation, \textbf{we need mixed half and double precision}: the dose deposition matrices are stored in half precision, while the input and output vectors are in double precision. The matrix is stored in half precision to save space, while the input and output vectors (which are the spot weights and output dose, respectively) are in double precision to ensure the stability of the optimization algorithm used in RayStation. While cuSPARSE supports mixed precision, the specific case of half and double precision mixing is not currently supported.

\begin{lstlisting}[
    language=C++,
    keywordstyle=\color{blue}\ttfamily,
    stringstyle=\color{red}\ttfamily,
    commentstyle=\color{green}\ttfamily,
    numbers=left,
    stepnumber=1,
    basicstyle=\tiny, %or \small or \footnotesize etc.
    caption={Example CUDA code for our CSR SpMV kernel. This example code is given in single precision only for clarity, but can be made to support mixed precision using C++-templates.},
    captionpos=b,
    label=spmv,
    float
]

__global__
 void CSR_SpMV(const float* __restrict__ data,
               const int* __restrict__ col_inds,
               const int* __restrict__ row_ptr,
               const float* __restrict__ x,
               float* __restrict__ y,
               int y_size)
{
    namespace cg = cooperative_groups;
    cg::thread_block block_group = cg::this_thread_block();
    cg::grid_group grid = cg::this_grid();

    int tile = cg::tiled_partition<warp_size>(block_group);
    int warp = block_group.group_index().x * tile.meta_group_size()
               + tile.meta_group_rank();
    int num_warps = grid.size() / warp_size;
    int idx_in_warp = tile.thread_rank();

    for (int i = warp; i < y_size; i += num_warps) {
        int start_row = row_ptr[i];
        int end_row = row_ptr[i + 1];

        if (start_row == end_row)
            continue;

        float val = 0.0;
        for (int j = start_row + idx_in_warp; j < end_row; j += tile.size())
            val += data[j] * x[col_inds[j]];

        float warp_aggregate = cg::reduce(tile, val, cg::plus<float>());

        if (idx_in_warp == 0)
            y[i] = warp_aggregate;
    }
}
\end{lstlisting}

\section{Experimental Setup}
\label{experiment}
In this work, we export the dose deposition matrices used from RayStation. The export is carried out directly in the RayStation source code, after the dose deposition matrix has been calculated by RayStation's Monte Carlo dose engine and before the optimization of the treatment plan. Recall that the dose deposition matrix does not change during the optimization process, since it reflects the physics of how the proton beam deposits dose in the patient. We then convert the dose deposition matrices from RayStation's in-house storage format to the CSR format used in the experiments.
To evaluate our implementation against the performance of the current RayStation algorithm, we measure the performance of the RayStation CPU implementation, which is what is used clinically in RayStation at the time of writing. Since having only this comparison might be misleading, due to our implementation being run on GPU instead of CPU, we also port the RayStation algorithm to GPU. We will refer to this implementation as \texttt{GPU Baseline} in the remainder of this paper. To note is that the RayStation CPU implementation does not map very well to GPU, mainly due to the use of per-thread scratch arrays to avoid race conditions while maintaining reproducibility. Due to the much larger number of threads in use on GPUs, the use of per-thread scratch arrays is not feasible on GPU. To address this, we utilize atomic operations instead in our GPU implementation of the RayStation algorithm. However, the use of atomic operations has one major drawback, namely that it causes the \texttt{GPU Baseline} implementation to not produce bitwise reproducible results between runs on the same system. As such, the \texttt{GPU Baseline} implementation is intended to give an optimistic bound on the performance of the RayStation algorithm on GPU.

We evaluate our CUDA implementations on three systems with Nvidia A100, V100 and P100 GPUs. Furthermore, we evaluate the current RayStation CPU implementation on an Intel i9-7940X CPU.
\begin{itemize}
    \item The A100 GPUs used in this work have 40 GBs of RAM and an L2 cache size of 40 MB. The system used for the A100 tests are equipped with AMD EPYC 7302P CPUs. We compile our code for the A100 using NVCC and CUDA version 11.1, with GCC 8.3.1 as the host compiler.
    \item Our V100 experiments are run on the Kebnekaise supercomputing cluster. The GPU nodes we used at Kebnekaise are equipped with two Nvidia V100 GPUs, with 16 GBs of RAM and 6 MB of L2 cache each, and two Intel Xeon Gold 6132 processors. We compile our code for the A100 using NVCC and CUDA version 10.1, with GCC 8.3.0 as the host compiler.
    \item For the P100 GPU, we compile our code using CUDA version 10.1 and GCC 4.8.5 as the host compiler. The P100 GPU has 16 GBs of RAM and 4 MB of L2 cache. The system the P100 experiments are run on are equipped with IBM POWER8 CPUs.
    \item The RayStation CPU tests were run on an Intel i9-7940x CPU with 64 GBs of RAM. The tests were done on a research version of RayStation 10B.
\end{itemize}

We test our implementation using half precision for the matrix elements, and double precision in the input and output vectors, as per the requirements in RayStation. Furthermore, to assess the performance of our implementation, we compare against existing library implementations of SpMV for CSR matrices, namely Ginkgo and cuSPARSE. Since these libraries do not, at the time of writing, support mixed double and half precision in SpMV, this comparison is done in single precision only. While this is not the precision used in RayStation, the comparison can be used to assess the performance of our implementation, as well as investigate the potential impact of different floating point precisions on performance. Our kernel using mixed half and double precision will be referred to as \texttt{Half/double}, and our implementation using single precision only will be referred to as \texttt{Single}.

Furthermore, we perform a roofline~\cite{williams2009roofline} analysis of our implementation, as well as of cuSPARSE and Ginkgo on the A100 GPU system. We use Nvidia's Nsight Compute to measure the total size of all memory transactions from DRAM to the caches. Given the total number of floating point operations required for SpMV, namely two times the number of non-zeros of the matrix, we calculate the arithmetic intensity of the SpMV implementations, which we then use in our roofline analysis.

We repeat our experiments 10000 times each, the values presented in the results section are the averages of those runs. We omit errorbars in the results in cases where the standard deviation is less than 5\%.

\section{Results}
\label{results}
When developing a new implementation of a particular algorithm, as is the case in the present paper, it is imperative to analyze how far said implementation is from the theoretical peak performance bound by machine characteristics. We apply the Roofline methodology in order to quantify how much of system compute- and memory-bandwidth resources we use. We perform this analysis on the A100 by analyzing the workloads and their arithmetic intensity (Flops/Byte) and empirically measuring their last-level cache (LLC) to external memory (HBM2 or DDR) using performance counters. We apply said methodology to all implementations, even those which currently cannot be adapted to RayStation (cuSPARSE and Ginkgo), to contrast our implementation against state-of-the-art. We do this for a sub-set of the liver and prostate use-cases.

\begin{figure}[t]
    \centering
    \includegraphics[width=\linewidth]{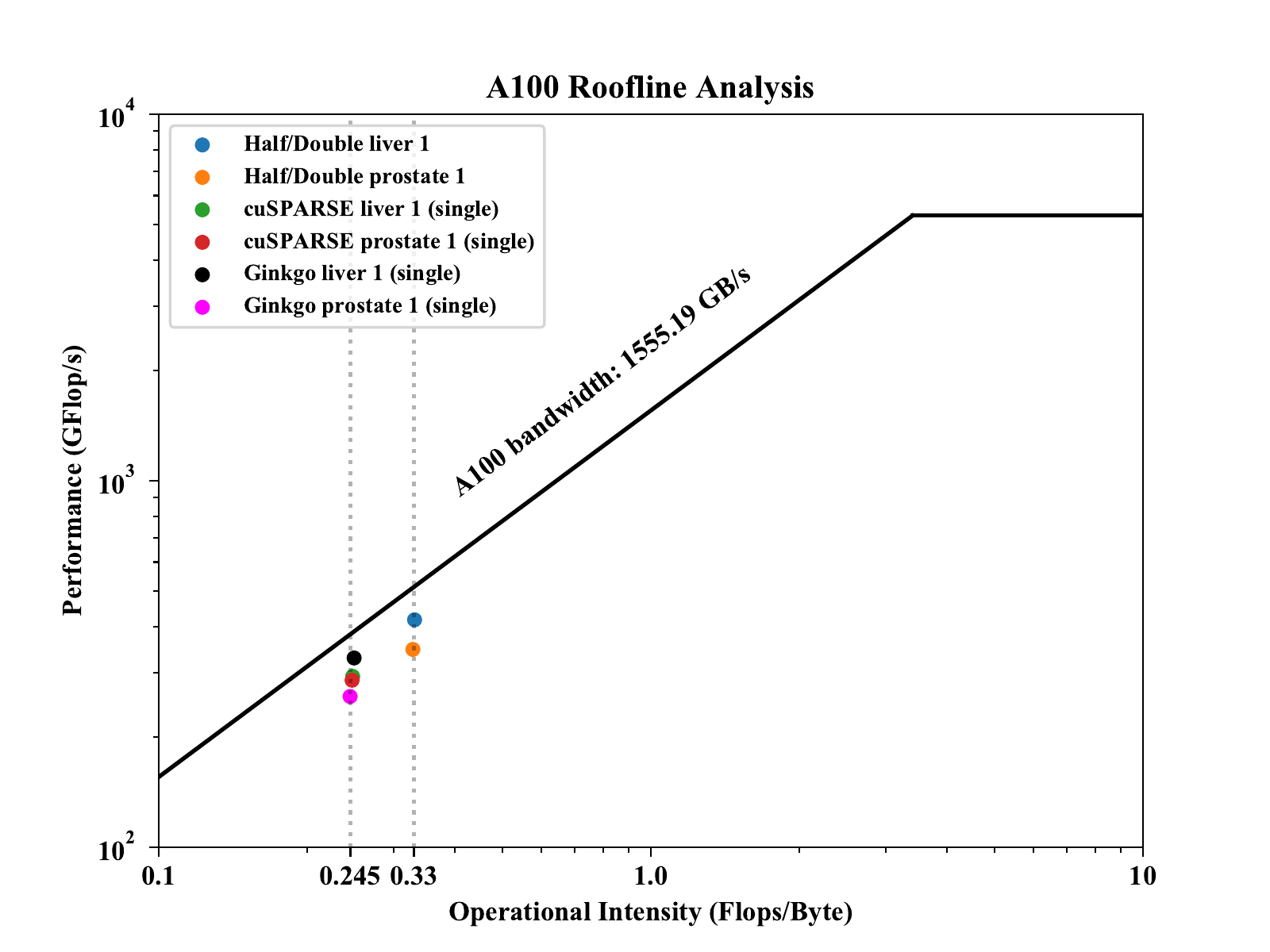}
    \caption{Roofline analysis of the Ginkgo SpMV kernel, cuSPARSE kernel and our mixed half and double precision implementations.}
    \label{fig:roofline}
\end{figure}

Figure~~\ref{fig:roofline} shows the results for the A100, which is the latest generation Nvidia GPU. In this plot, the performance (y-axis) is a function of the operational intensity (x-axis). We measure the operational intensities used in Figure~\ref{fig:roofline} the \texttt{dram\_\_bytes} metric from Nvidia's Nsight Compute. However, in the case of SpMV in particular, we note that the operational intensity can be calculated accurately using a simple model as well, even when accounting for cache filtering. To illustrate this, we compute a theoretical upper bound estimate for the operational intensity for liver~beam~1.

To obtain an upper bound on the operational intensity for the \texttt{Half/Double} case, we assume an infinite cache size, such that all values need only be read from main memory once, and then can be accessed from cache. In the case of CSR SpMV, for each multiplication and addition we perform, we need to load one 2 byte value from the matrix, together with a 4 byte column index, and one 8 byte value from the input vector. Furthermore, for each row of the matrix that we process, we load its starting and final index in the flattened data array from the \texttt{row\_ptr} array (see lines 21 and 22 in Listing \ref{spmv}). Since the \texttt{end\_row} is the \texttt{start\_row} index for the next row (and assuming infinite cache size) we only need to load one index of four bytes per row of the matrix from main memory, except for the first row, where both indices must be loaded from main memory. Finally, we need to write our local aggregate value for each row (corresponding to the dot product between the row and the input vector) to the output vector. For the sake of estimating an upper bound, let's assume that the output vector has non-zero values in all positions, meaning that we need to write 8 bytes for each row in the matrix to the output vector. The total memory traffic required then becomes $6 * nnz + 12 * nr + 8 * nc$, where $nnz$ is the number of non-zeros in the matrix, $nr$ is the number of rows and $nc$ the number of columns. Given that the number of floating point operations required is $2 * nnz$, we can compute the operational intensity for liver~beam~1 using the values from Table~\ref{tab:dimensions}. This gives us an approximation of the upper bound for the operational intensity of $0.332$, which is very close to the measured value seen in Figure~\ref{fig:roofline}.

The reason the measured and theoretical peak operational intensity are so similar in value likely comes from the total memory transactions required being dominated by the number of non-zeros in the input matrix, which is several orders of magnitude larger than the number of rows and columns. Since each non-zero in the matrix is only used once in the calculation, the infinite cache re-use assumption has no effect, since no re-use is required. Furthermore, the dimensions of the input vector, whose values are re-used, are small enough to fit entirely in the 40MB L2 cache of the A100. This also explains the difference in operational intensity between the \texttt{Single} version and \texttt{Half/double}. Since the memory traffic required is dominated by the non-zeros of the matrix, increasing the size of the matrix values by 2 bytes has a noticeable impact on the observed (and theoretical) operational intensity. This higher operational intensity for the \texttt{Half/Double} case, when compared to the single precision only experiments, like for Ginkgo and cuSparse, leads to better bandwidth utilization and thus higher performance.

Finally, the analysis of the operational intensity also reveals another important factor in the performance of the kernel, namely that the memory traffic caused by loading the column indices of the elements in the matrix make up a large portion of the total memory traffic. Assuming that 4 byte integers are used to store the column indices (which is the case in this work), the column indices then contribute $4 * nnz$ bytes to the total memory traffic. Since the $nnz$ term typically dominate the others in the formula for the total memory traffic, the contribution to the memory traffic by the column indices is substantial. We can already see in figure \ref{fig:roofline} that a reduction in the size of the matrix elements from 4 bytes to 2 bytes, effectively reducing the memory traffic contribution from them by $2 * nnz$, has a significant impact on the operational intensity. It is reasonable to expect that the same reduction in the size of the column indices would similarly increase the operational intensity. Looking at the dimensions from table \ref{tab:dimensions}, the column indices for the prostate case could be stored using 16 bit unsigned integers, thus saving memory and likely improving performance. While the column indices for the liver cases cannot be directly stored using 16-bit integers, they are not much larger than the largest possible value for 16 bit integers, namely 65535. Through this analysis, we can identify the size used to store column indices in memory as a potential improvement for future work, as the matrices arising from the specific case of proton dose calculation have a significantly smaller number of columns compared to rows.

In summary, our implementation, using mixed half and double precision, increases the amount of operations we do per external memory access; improvement in this metric is particularly important in GPUs, which often are bandwidth-deprived machines (their Flops/Byte ratio is rather high). As such, we expect the performance of the kernels using mixed half and double precision to perform better than the ones using only single precision. SpMV is typically a highly memory-bound operation, and this is true in this case as well, as can be seen from the low operational intensity of the evaluated kernels. Due to this, we consider also the bandwidth achieved when evaluating the performance later in section \ref{sec:perf_measurements}.
\subsection{Execution Configuration}
To determine the most performant execution configuration (number of threads per block and thread blocks) for our kernels, we perform some experiments with varying execution configurations to determine the best one. In our CSR SpMV kernel, since we assign one thread warp of 32 threads to each row, the total number of threads we use in our execution configuration is 32 times the number of rows of the dose deposition matrix. To further narrow down a more specific number of thread blocks and threads per block to use, we perform tests using a varying number of threads per block for liver~beam~1 and measure the performance in order to find the best configuration. In the experiments, we vary the number of threads per block between 32 and 1024, while setting the number of thread blocks such that the total number of threads is 32 times the number of rows.

\begin{figure}[t]
    \centering
    \includegraphics[width=\linewidth]{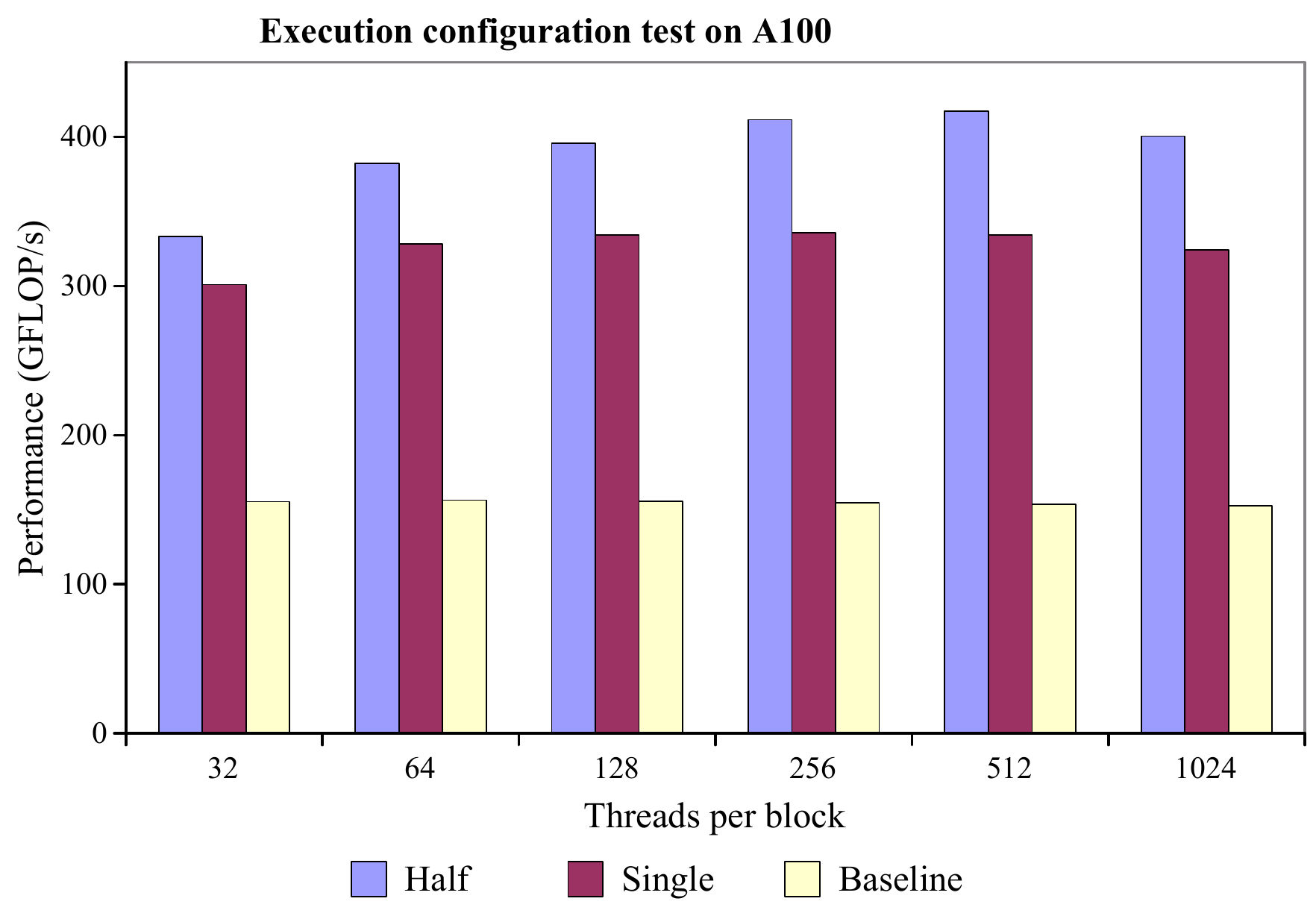}
    \caption{Performance for our different kernels with varying number of threads per block on liver~beam~1. The total number of threads used is fixed by the input size, so the number of thread blocks is always set such that the total number of threads is 32 times the number of rows.}
    \label{fig:exec_config}
\end{figure}

From inspecting Figure~\ref{fig:exec_config}, it is clear that 512 threads per block performs the best for the \texttt{Half/Double} implementation on liver~beam~1. For the \texttt{Single} kernel, the performance is similar for 128, 256 and 512 threads per block. Finally, for the \texttt{Baseline} kernel, the performance is also similar for different execution configurations, with 64 and 128 threads per block performing  slightly better. In the following experiments, we use 512 threads per block for \texttt{Half/Double} and \texttt{Single} and 128 threads per block for \texttt{Baseline}.


\subsection{Performance Measurements} \label{sec:perf_measurements}
\begin{figure}[t]
    \centering
    \includegraphics[width=\linewidth]{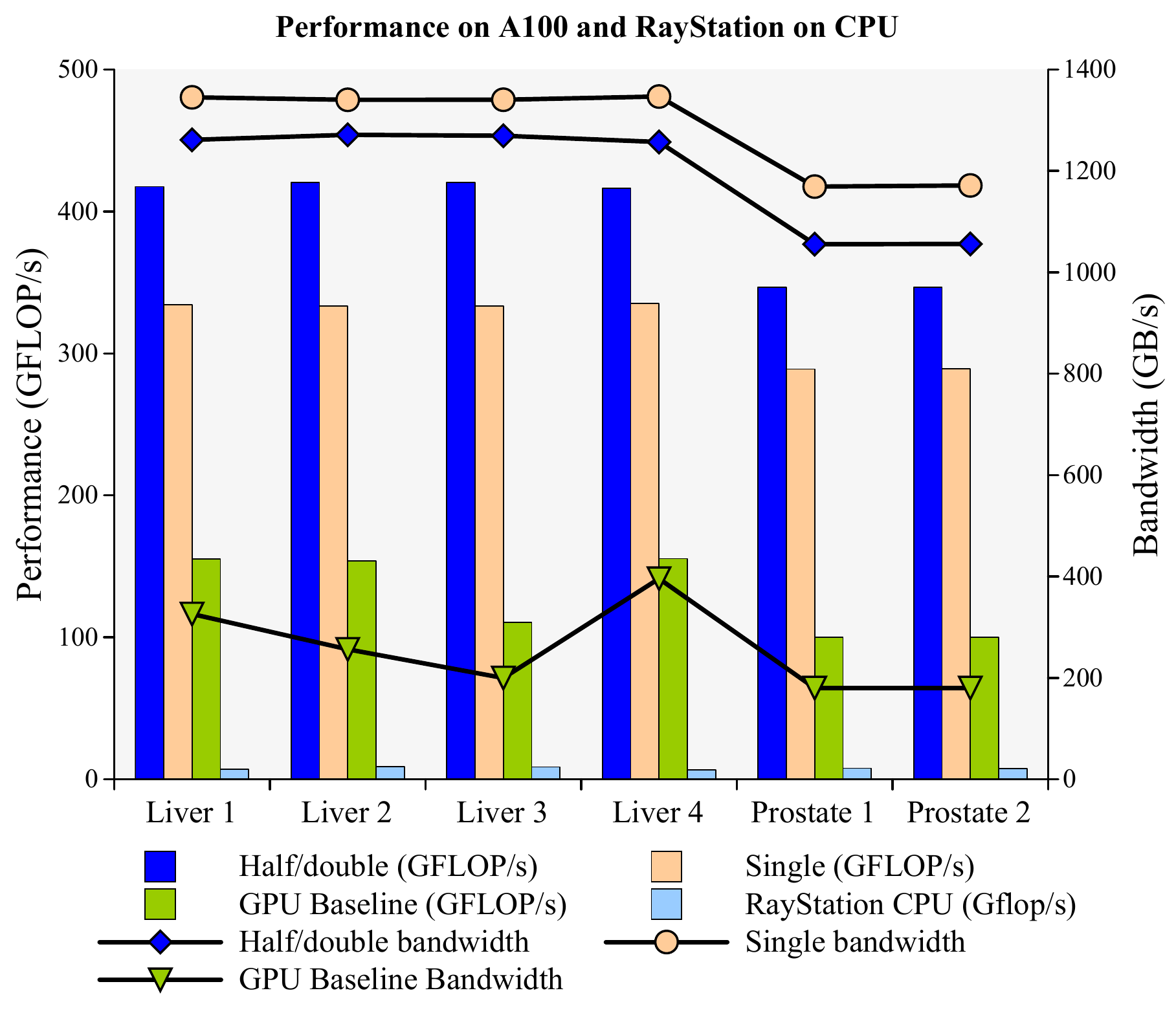}
    \caption{Performance measurements of our implementations on the A100 and RayStation's CPU implementation. The GPU baseline is the current algorithm used in RayStation ported to GPU. Half/double and Single precision are our implementations with mixed half and double precision and single precision only, respectively. The right y-axis shows the bandwidth values, and the left axis the GFLOP/s. The lines with markers show the bandwidth values and the bars the GFLOP/s.}
    \label{fig:A100_perf}
\end{figure}
Figure~\ref{fig:A100_perf} shows the performance of the different implementations. Here, the units of metric are the floating-point performance (FLOP/s) and bandwidth. The bandwidth is measured using performance counters from Nvidia Nsight Compute, with the measurement being done for memory traffic between the L2 cache and DRAM. We test the different implementations against different use cases with different dose deposition matrices (4 for liver cancer and 2 for prostate cancer). We compare three different versions. \texttt{Baseline} is the direct port of the algorithm to the GPU based on RayStation's CPU code (also using RayStation's custom compressed format). \texttt{Half/Double} is our optimized implementation (our contribution in the paper). The \texttt{Single} implementation provides a contrast against the other two, as well as for comparison with popular library implementations (using single precision for the input).

Out of the three GPU implementations, \texttt{GPU Baseline} is the poorest performing. This is expected as the CPU algorithm does not map well to the GPU, mainly due to the use of scratch arrays to avoid race conditions in the CPU implementation. We address this issue in the GPU implementation by using CUDA atomics instead, which incurs additional overhead. The implementation we engineer for this work performs rather well-- up-to 4x faster than the \texttt{GPU Baseline} for liver~beam~3 with an average speedup of approximately 3$\times$. The highest performance we reach is 420 GFlop/s, which is $\Tilde{}8\%$ of peak Nvidia A100 double precision floating point performance, on the liver use-cases. We also note that there is a rather large difference between cancer treatment depending on the use-case, where the liver use-cases often experience a ~30\% improvement over prostate. This degradation of performance materialize could be caused by the relatively smaller size of the prostate cases, with significantly smaller matrices. The \texttt{Single} version performs worse than the \texttt{Half/double}. The reason behind this is likely that the \texttt{Half/double} version have a higher arithmetic intensity (Flops/Byte) than the single-precision. This is discussed further in the beginning of this section. The RayStation CPU implementation is noticeably slower than the GPU implementations on the A100. A large part of this can be explained by the difference in hardware used, since the GPU port of the RayStation code already shows a $~17 \times$ speedup when compared to the CPU implementation. In practice, this indicates that significant performance benefits could be achieved by moving the implementation of the SpMV for dose calculation to GPU, which in turn could improve optimization times in RayStation significantly for proton plans.

\begin{figure}[t]
    \centering
    \includegraphics[width=\linewidth]{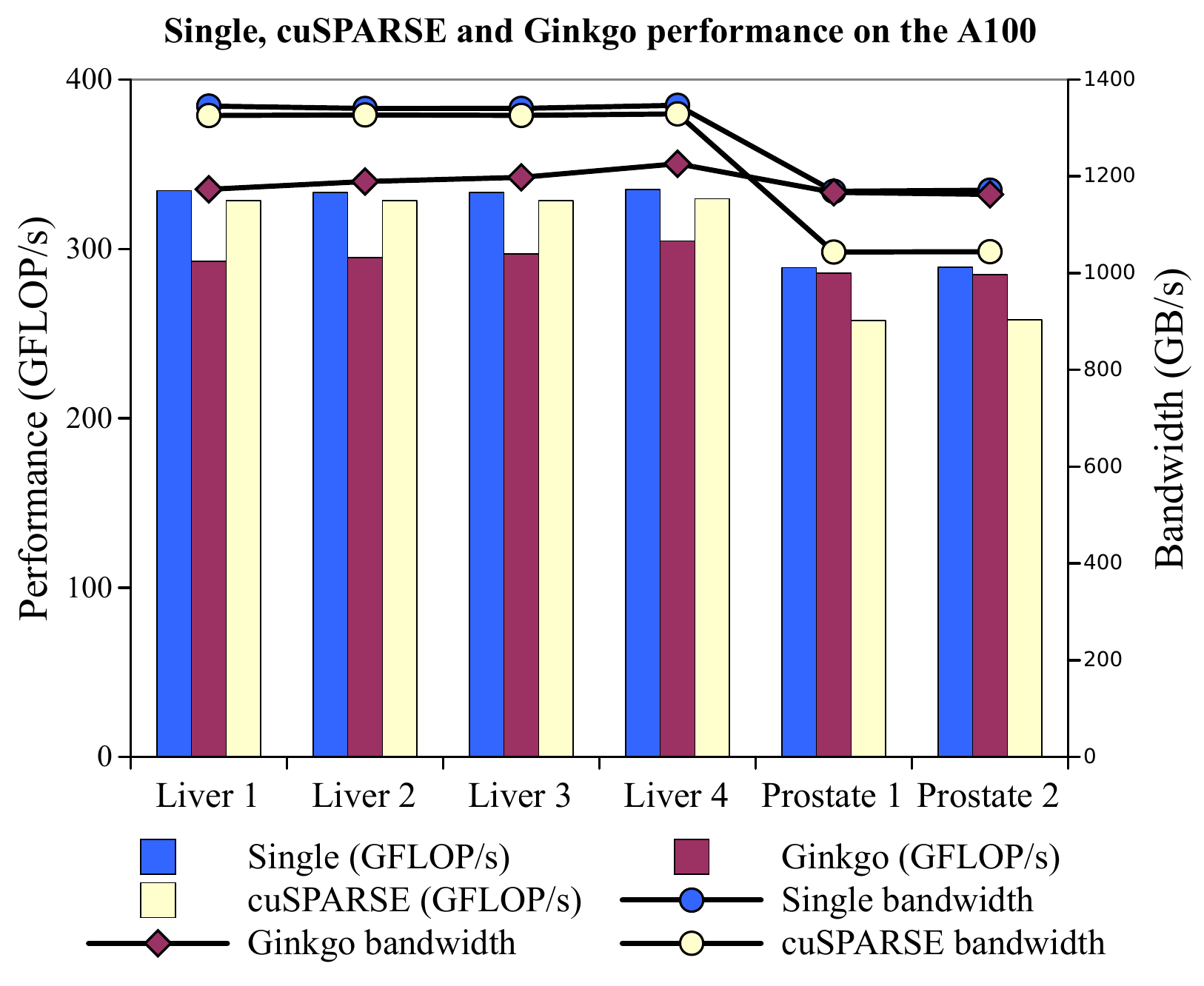}
    \caption{Comparison of performance between our kernel, cuSPARSE and Ginkgo in single precision on the A100. The right y-axis shows the bandwidth values, and the left axis the GFLOP/s. The lines with markers show the bandwidth values and the bars the GFLOP/s.}
    \label{fig:library_perf}
\end{figure}
In terms of bandwidth, the \texttt{Half/double} and \texttt{Single} kernels have a bandwidth of about ~1250-1350 GB/s on the liver cases, which is approximately $80-87 \,\%$ of the peak bandwidth of 1555 GB/s for the A100. The bandwidth on the prostate cases is slightly lower at approximately 1050 GB/s, or about $68 \,\%$ of the theoretical peak bandwidth. As in the case for the GFlop/s, the bandwidth achieved for the prostate cases is slightly lower than for the liver cases, possibly due to smaller matrix sizes. For the \texttt{GPU Baseline}, we see that the measured bandwidth varies significantly between the different cases. One possible explanation for this is that the bandwidth is measured from the L2 cache to DRAM. The \texttt{GPU Baseline} kernel, due to using atomic reductions to the output vector, write to the output vector much more frequently compared to the \texttt{Half/double} and \texttt{Single} kernels, that only write to the output vector once per row. This significantly increases the total memory traffic for the \texttt{GPU Baseline} kernel. However, since the 40 MB L2 cache of the A100 is large enough for the entire output vector to fit in, this additional memory traffic is mostly between caches. This means the kernel overall is more constrained by memory traffic inside the caches, meaning that accesses to DRAM for other required data like matrix elements happen less frequently, causing lower overall bandwidth and possibly a higher variation in the DRAM bandwidth achieved, as seen in Figure~\ref{fig:A100_perf}.

Furthermore, Figure~\ref{fig:library_perf} shows a performance comparison between our \texttt{Single} kernel, Ginkgo and cuSPARSE. We see that our implementation matches the performance or is better than Ginkgo and cuSPARSE for the evaluated matrices. When comparing cuSPARSE to Ginkgo, we observe that cuSPARSE performs better than Ginkgo for the liver cases but worse for the prostate cases. Since the implementation details of cuSPARSE are not public, the reason for this difference is difficult to give. Overall, we observe that our implementation's performance is comparable or better compared to state-of-the-art library implementations for SpMV, when the computation is done in single precision only. Looking at the bandwidth values from Figure~\ref{fig:library_perf}, we see that it follows the performance trends noted in the FLOP/s very closely. This is somewhat expected given the highly memory-bound nature of SpMV.

\begin{figure}[t]
    \centering
    \includegraphics[width=\linewidth]{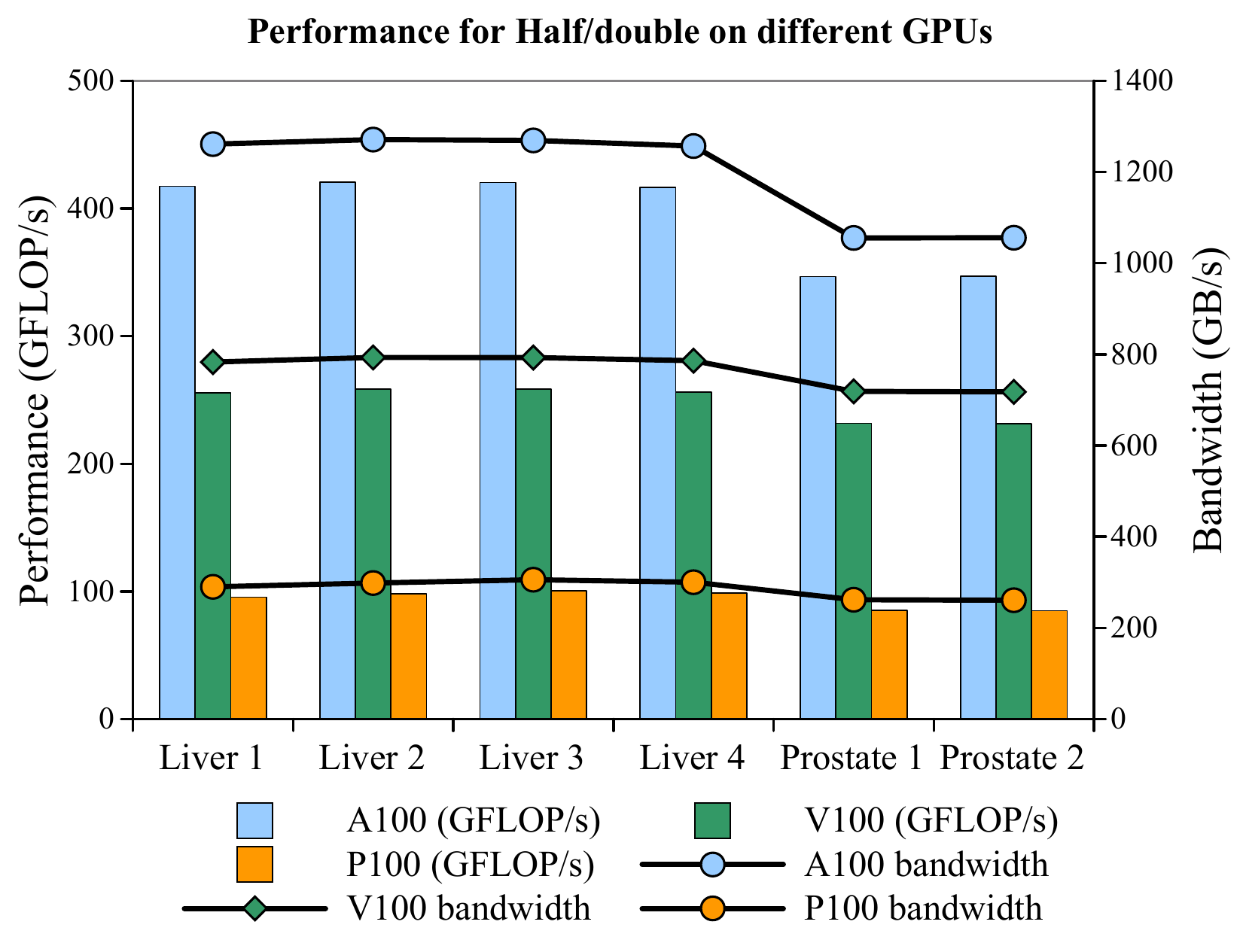}
    \caption{Performance measurements on the A100, V100 and P100 for the \texttt{half/double} implementation. The right y-axis shows the bandwidth values, and the left axis the GFLOP/s. The lines with markers show the bandwidth values and the bars the GFLOP/s.}
    \label{fig:machine_comp}
\end{figure}

Finally, Figure~\ref{fig:machine_comp} shows a comparison between the performance for the A100, V100 and P100 systems. The difference in performance between the A100 and V100 is approximately between a factor 1.5 and 2. The difference in performance between the V100 and P100 is larger, with a difference of a factor of 2.5, approximately. This difference in performance cannot be fully explained by the difference in peak memory bandwidth of the GPUs, which are 1555 GB/s for the A100, 897 GB/s for the V100 and 732 GB/s on the P100, even though SpMV is typically a memory-bound kernel. On both the A100 and V100 we attain roughly 80-88\% of the peak bandwidth. On the P100, the achieved value is closer to 41\%, a large difference. We intend to investigate the causes of the difference in performance between the different GPUs in future work.

\section{Related Work}
\label{related_work}
Several previous works have focused on accelerating medical treatment planning using GPUs. Applications in the treatment planning workflow that have been accelerated using GPUs previously include: algorithms for image processing and deformable registration, proton planning using Monte-Carlo simulation on GPUs~\cite{qin2016recent, schiavi2017fred, souris2016fast}, dose calculation for different treatment modalities as well as the optimization process itself. An overview of areas in RTP where GPU acceleration has been used can be found in the work by Jia et al.~\cite{jia2014gpu}. A condensed history and overview of challenges on proton planning can be found at~\cite{mohan2017proton}.

There is a rich lineage of research literature on Sparse Vector-Multiplication, tracing back all the way to the 1970s \cite{duff1977survey}, and we would do a poor job condensing it within this paper. Instead, we redirect the interested reader to several well-written surveys  on the topic~\cite{mcssc07, gpusc09, gpusc14, grossman2016survey}.

\section{Discussion and Conclusion}
\label{conclusions}
In this paper, we described the porting of radiation dose calculation, a major computational bottleneck in RTP, to Nvidia GPUS. We ported the original CPU implementation used in the RaySearch engine to GPU and found several opportunities for optimization. Our contributed implementation, which leverages mixed-precision in a way that is not allowed yet in state-of-the-art SpMV implementations such as cuSPARSE and Ginkgo, improves performance by up to 4$\times$ (average: 3$\times$) over the baseline and have a higher operational intensity compared to both cuSPARSE and Ginkgo. Furthermore, the GPU port of RayStation's CPU algorithm shows an approximate $17 \times$ performance improvement compared to the performance on the CPU. With our modified CSR kernel, the performance improvement is even larger at $46 \times$ improvement compared to the CPU baseline. In practice, this can mean a significant speedup in optimization times and time-to-treatment for radiation therapy treatment planning.

We believe that our results encourage the wider inclusion of high-performance accelerators into medical facilities, as they will reduce the time-to-treatment for cancer patients and enable the use of more sophisticated RTP methods, ultimately improving care for cancer patients. In the future, together with the investigation of sparse matrix formats (other than CSR), we intend to investigate other components of the Radiation Therapy pipeline in order to research and further reduce time-to-treatment.



\section*{Acknowledgments}
The computations were performed on resources provided by the Swedish National Infrastructure for Computing (SNIC) at High Performance Computing Center North (HPC2N) partially funded by the Swedish Research Council through grant agreement no. 2018-05973.



\bibliographystyle{IEEEtran}
\bibliography{IEEEabrv,References}
%



\end{document}